\documentclass[aps,superscriptaddress,twocolumn,prx]{revtex4-1}
\usepackage{amssymb,amsmath,mathptmx}
\usepackage{graphicx}% Include figure files4T_3.fig
\usepackage{dcolumn}% Align table columns on decimal point
\usepackage{bm}% bold math
\usepackage{hyperref}% add hypertext capabilities
\usepackage{color}
\usepackage[utf8x]{inputenc}%french
\usepackage{array}
\usepackage{ulem}
\newcolumntype{C}{>{\centering\arraybackslash}p{1em}}

\newcommand{\be}{\begin{equation}}
\newcommand{\ee}{\end{equation}}
\newcommand{\bea}{\begin{eqnarray}}
\newcommand{\eea}{\end{eqnarray}}
\begin{document}

\title{Magnetointerferometry of multiterminal Josephson
  junctions}

\author{R\'egis M\'elin}
\email{regis.melin@neel.cnrs.fr}

\affiliation{Université Grenoble-Alpes, CNRS, Grenoble INP\thanks{Institute
    of Melin2019 Univ. Grenoble Alpes}, Institut NEEL,
  Grenoble, France}

\author{Clemens B. Winkelmann}

\affiliation{Université Grenoble-Alpes, CEA, Grenoble INP\thanks{Institute
    of Melin2019 Univ. Grenoble Alpes}, IRIG-Pheliqs, Grenoble, France}

\author{Romain Danneau}

\affiliation{Institute for Quantum Materials and Technologies,
  Karlsruhe Institute of Technology, Karlsruhe D-76021, Germany}

\begin{abstract}
  We report a theoretical study of multiterminal Josephson junctions
  under the influence of a magnetic field $B$.  We consider a
  ballistic rectangular two-dimensional metal $N_0$ connected by the
  edges to the left, right, top and bottom superconductors $S_L$,
  $S_R$, $S_T$ and $S_B$, respectively.  We numerically calculate in
  the large-gap approximation the critical current
  $I_c$ versus $B$ between the left and right $S_L$
  and $S_R$ for various aspect ratios, with the top
    and bottom $S_T$ and $S_B$ playing the role of superconducting
  mirrors. We find the critical current $I_c$ to be
  enhanced by orders of magnitude, especially at long distance, due to
  the phase rigidity provided by the mirrors. We obtain
  superconducting quantum interference device-like magnetic
  oscillations. With symmetric couplings, the self-consistent
  superconducting phase variables of the top and bottom mirrors take
  the values $0$ or $\pi$, as for emerging Ising degrees of
  freedom. We propose a simple effective Josephson junction circuit
  model that is compatible with these microscopic numerical
  calculations. From the $I_c(B)$ patterns we infer where the
  supercurrent flows in various device geometries.  In particular in
  the elongated geometry, we show that the supercurrent flows between
  all pairs of contacts, which allows exploring the full phase space
  of the relevant phase differences.
\end{abstract}

\maketitle

\section{Introduction}

Superconducting multiterminal systems have recently attracted
considerable attention. While early theoretical works already
predicted unusual behavior of these more complex Josephson
junctions~\cite{Ouboter1995,Amin2001,Amin2002,Amin2002a}, the later
ones demonstrated that these systems may host several exotic phenomena
such as correlations among Cooper pairs known as the
quartets~\cite{Freyn2011,Melin2016,Jonckheere2013,
  Melin2017,Melin2019,Doucot2020,Melin2020,Melin2020a,Melin2021,Melin2022,Melin2023,Melin2023a,Keleri2023}
as well as Weyl points singularities and nontrivial topology in the
Andreev bound state
spectrum~\cite{Riwar2016,Eriksson2017,Xie2017,Xie2018,Deb2018,Venitucci2018,Gavensky2018,Klees2020,Fatemi2021,Peyruchat2021,Weisbrich2021,
  Chen2021,Chen2021a,Repin2022,Gavensky2023}, and the energy level
repulsion in Andreev
molecules~\cite{Pillet2019,Pillet2020,Kornich2019,Pillet2023}.
Following these theoretical efforts, recent experiments have reported
the detection of Cooper
quartets~\cite{Pfeffer2014,Cohen2018,Huang2022,Graziano2022}, the
observation of Floquet-Andreev states~\cite{Park2022}, the studies of
Andreev
molecules~\cite{Kurtossy2021,Coraiola2023,Matsuo2023,Matsuo2023a}, the
multiterminal superconducting diode effect~\cite{Gupta2023,Zhang2023},
in addition to other results using numerous different types of
superconducting weak links
~\cite{Matsuo2022,Draelos2019,Pankratova2020,Graziano2020,Arnault2021,Arnault2021,Khan2021,Arnault2022,Zhang2022}.

The common ground to these models and experiments is related to the
fact that the weak links are connected by, at least, three
superconducting contacts. Indeed, in comparison to its two leads
counterparts, the supercurrent flow in multiterminal Josephson
junctions may appear non-trivial. Seminal works showed that the
supercurrent distribution could be probed by analyzing the
interference pattern induced by the application of a magnetic flux
across two-terminal Josephson junctions
\cite{Rowell1963,Dynes1971,Zappe1975,BaronePaternoBook,Tinkham}. Therefore,
this interferometric pattern strongly depends on the device geometry
and where the supercurrent flows
\cite{Barzykin1999,Kikuchi2000,Angers2008,Chiodi2012,Amado2013,Hart2014,Allen2016,Meier2016,Amet2016,Kraft2018,Irfan2018,Pandey2022,Chu2023}.
As shown by Dynes and Fulton~\cite{Dynes1971}, in two-terminal
Josephson junctions, the magnetic field dependence of the critical
current is related to the supercurrent density distribution across the
device by an inverse Fourier transform as long as the supercurrent
density is constant along the current flow. However, alternative
models are needed in the case of non-homogeneous supercurrent density
\cite{Kraft2018} or non-regular shapes \cite{Irfan2018,Chu2023}.

\begin{figure*}[htb]
  \begin{minipage}{.59\textwidth}
    \centerline{\includegraphics[width=\textwidth]{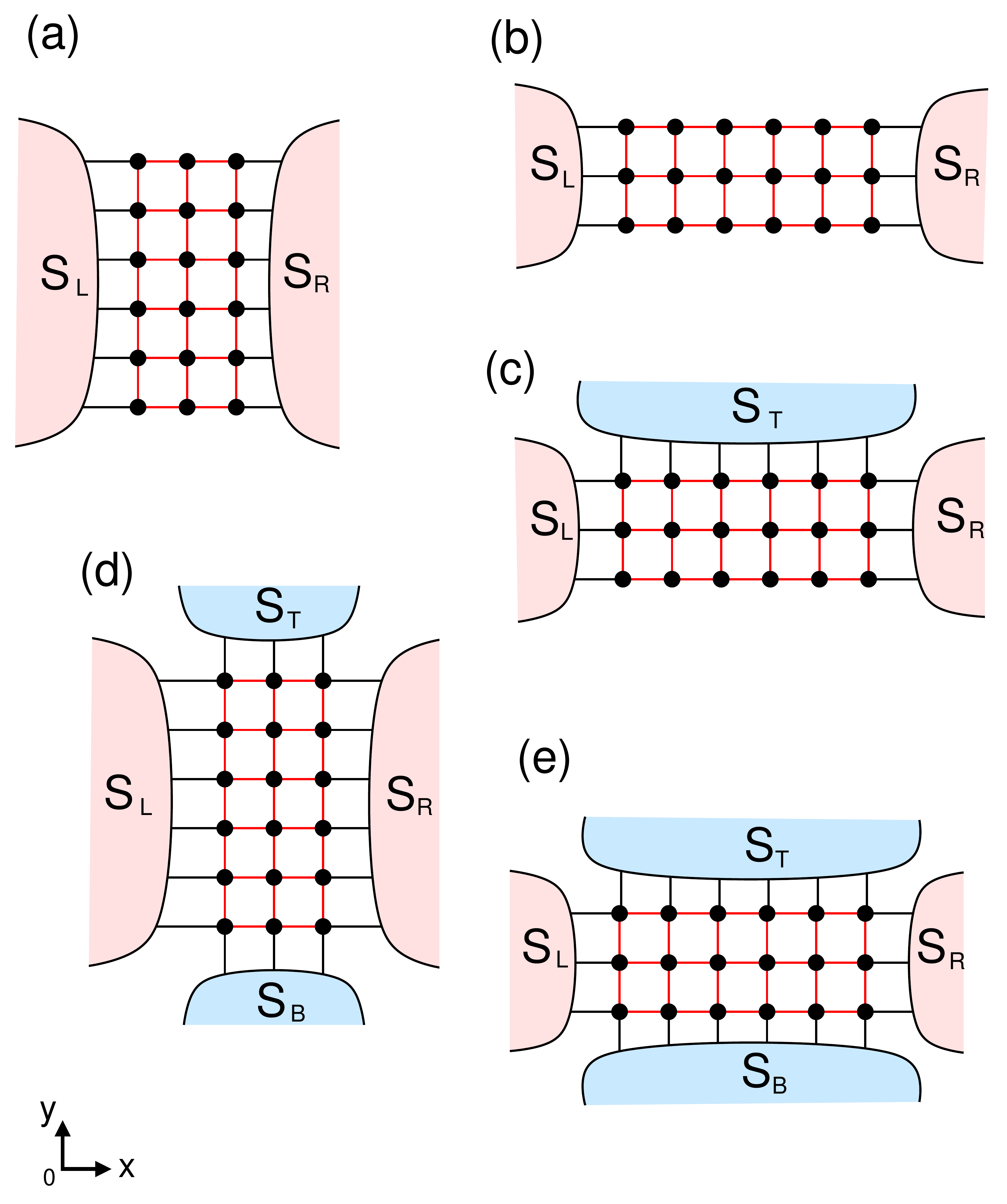}}\end{minipage}\begin{minipage}{.4\textwidth}
    \caption{Schematics of the considered two- and multiterminal
      Josephson geometries. The superconductors $S_T$ and $S_B$ on top
      and bottom are in open circuit, that is, they are
      superconducting mirrors, and we calculate the current flowing
      horizontally from $S_L$ to $S_R$. A four-terminal device with
      $N=3$ and $M=6$ is shown on panel a, i.e. a device with $M/N\agt
      1$ elongated in the vertical $y$-axis direction. A two-terminal
      device elongated in the horizontal $x$-axis direction is shown
      on panel b with $N=6$ and $M=3$, i.e. with $M/N\alt 1$. Panels
      c, d and e feature three- or four-terminal devices containing a
      single or two superconducting mirrors, and elongated along the
      $x$- or $y$-axis directions.
      \label{fig:the-device}
  }\end{minipage}
\end{figure*}

To our knowledge, no theories exploring the current flow and the
corresponding magneto-interferometric pattern in multi-terminal
Josephson junctions are available so far. Here, we present a
microscopic model allowing us to calculate the magnetic field
dependence of the critical current in various configurations (see
Fig.~\ref{fig:the-device}).  Our calculations are based on a large-gap
Hamiltonian in which the supercurrent is triggered by the {\it tracer}
of the phase of the vector potential, i.e. we calculate the critical
current pattern as a function of the magnetic field. While we recover
the standard two-terminal interferometric patterns, we show that the
additional lead drastically modifies the magnetic field dependence of
the critical current. With four terminals, our calculations reveal
that the supercurrent visits all of the superconducting leads, which
could result from a kind of ergodicity. This notion of ergodicity was
lately pointed out via the studies of the critical current contours
(CCCs) in four-terminal Josephson junctions, as a function of two
different biasing
currents~\cite{Pankratova2020,Melin2023a}. Consistency was
demonstrated \cite{Pankratova2020} between the experiments on the CCCs
and Random Matrix Theory, where the scattering matrix bridges all of
the superconducting leads. Considering disorder in the short-junction
limit, quantum chaos leads to ergodicity in the sense of Andreev bound
states (ABS) coupling all of the superconducting leads. The
supercurrent significantly visits all of those {$n$
  superconducting terminals}, thus being sensitive to $n-1$
independent phase differences, a number that is however reduced by the
additional constraints of current conservation imposed by the external
sources. In the other limit of large-scale devices, another recent
work~\cite{Melo2022} pointed out the relevance of long-range effects
in multiterminal configurations, as the result of the phase rigidity.

Here, we also find long-range propagation of the supercurrent in
three- or four-terminal geometry having one or two superconducting
mirrors respectively, due to the phase rigidity in the leads under
zero-current bias condition. In the four-terminal geometries, the
leads $S_L$, $S_R$, $S_T$ and $S_B$ are connected to the left, right,
top and bottom sides of the rectangular normal-metallic conductor
$N_0$ and $S_T,\,S_B$ are laterally connected on top and bottom, being
superconducting mirrors in open circuit, as shown
Fig.~\ref{fig:the-device}c and Fig.~\ref{fig:the-device}d. For the
elongated geometry along the horizontal $x$-axis direction (see
Fig.~\ref{fig:the-device}c), the four-terminal magnetic oscillations
of the critical current resemble the pattern of a superconducting
quantum interference device (SQUID) because of the interfering
supercurrent paths propagating in $S_T$ and $S_B$ over long
distance. The critical current in the horizontal direction is
controlled by the phases $\varphi_T$ and $\varphi_B$ of the top and
bottom superconductors $S_T$ and $S_B$. Symmetry in the hopping
amplitudes connecting $N_0$ to the four superconductors leads to the
discrete values $\varphi_T,\,\varphi_B=0$ or $\pi$, as for emerging
Ising degrees of freedom.

Finally, a simple phenomenological Josephson junction circuit
model is proposed for devices elongated in the horizontal 
direction. In this model, both of the superconducting phase variables
$\varphi_T$ and $\varphi_B$ enter the critical current via their
difference $\varphi_T-\varphi_B$, which originates from the large
Josephson energy coming from the extended interfaces parallel to the
horizontal direction.

The paper is organized as follows. The model and Hamiltonians are
presented in Sec.~\ref{sec:themodel}. The numerical results are
presented and discussed in Sec.~\ref{sec:num}.
Sec.~\ref{sec:phenomenological} presents a phenomenological Josephson
junction circuit model. Concluding remarks are provided in
Sec.~\ref{sec:conclusions}.

%%%%%%%%%%%%%%%%%%%%%%%%%%%%%%%%%%%%%%%%%%%%%%%%%%%%%%%%%%%%%%%%%%%%%%%
%%%%%%%%%%%%%%%%%%%%%%%%%%%%%%%%%%%%%%%%%%%%%%%%%%%%%%%%%%%%%%%%%%%%%%%

\section{Model and Hamiltonians}
\label{sec:themodel}

In this section, we define the Hamiltonian of the devices shown in
Fig.~\ref{fig:the-device}. The Hamiltonians of each part of the
circuit are provided in subsection~\ref{sec:general}. The large-gap
Hamiltonian of the entire structure is presented in
subsection~\ref{sec:large-gap}, and the boundary conditions in the
presence of a magnetic field are next discussed in
subsection~\ref{sec:boundary}. The algorithm is presented in
Sec.~\ref{sec:algo}.

\subsection{General Hamiltonians}
\label{sec:general}

In this subsection, we introduce the Hamiltonians of the
superconductor, the central normal-metal conductor and the coupling
between them.

{\it The superconductors} are described by the BCS Hamiltonian
\begin{eqnarray}
  \label{eq:H-BCS1}
  \hat{\cal H}_{BCS}&=&-W \sum_{\langle i,j \rangle}
  \sum_{\sigma_z=\uparrow,\downarrow} \left(c_{i,\sigma_z}^+
  c_{j,\sigma_z}+ c_{j,\sigma_z}^+ c_{i,\sigma_z}\right) \\&-&
  \Delta \sum_k \left(\exp\left(i\varphi_k\right)
  c_{k,\uparrow}^+ c_{k,\downarrow}^+ +
  \exp\left(-i\varphi_k\right) c_{k,\downarrow}
  c_{k,\uparrow}\right) ,
  \label{eq:H-BCS2}
\end{eqnarray}
where the summation in the first term is over all pairs of neighboring
tight-binding sites $\langle i,j\rangle$ and over the projection
$\sigma_z$ on the spin quantization axis, that is the $z$-axis. The
first term given by Eq.~(\ref{eq:H-BCS1}) corresponds to the kinetic
energy, i.e. to spin-$\sigma_z$ electrons hopping between neighboring
tight-binding sites on a square-lattice. The second term given by
Eq.~(\ref{eq:H-BCS2}) is the mean field BCS pairing term, with
superconducting phase variable $\varphi_k$ at the tight-binding
site~$k$. The superconducting phase variables $\varphi_k$ take
different values between different superconducting leads and the
$\varphi_k$s are assumed to be uniform within each of those since we
handle weak currents throughout the paper. In order to reduce the
computational expanses, we carry out the calculations in a regime
where the superconducting gap is the largest energy scale, leading to
a {\it large-gap Hamiltonian} for the entire device connected to the
superconducting leads. This approach will be justified from
qualitative agreement with the known Fraunhofer pattern like in a
two-terminal configuration (i.e. with vanishingly small coupling to
the top and bottom $S_T$ and $S_B$ respectively, see
Figs.~\ref{fig:the-device}a and~\ref{fig:the-device}b).

{\it The central ballistic normal-metallic conductor} $N_0$ is
described by the square-lattice tight-binding Hamiltonian on a
rectangle of dimensions $N a_0\times M a_0$ in the horizontal $x$- and
vertical $y$-axis directions respectively, where $a_0$ is the lattice
spacing:
\begin{equation}
  \label{eq:H-Sigma-0}
  \hat{\cal H}_{\Sigma^{(0)}}=-\Sigma^{(0)} \sum_{\langle i,j \rangle}
  \sum_{\sigma_z=\uparrow,\downarrow} \left(c_{i,\sigma_z}^+
  c_{j,\sigma_z}+ c_{j,\sigma_z}^+ c_{i,\sigma_z}\right)
  ,
\end{equation}
with hopping amplitude $\Sigma^{(0)}$. Eq.~(\ref{eq:H-Sigma-0}) is
intended to qualitatively capture a two-dimensional conductor at high charge carrier density, and thus presenting a well-defined extended
Fermi surface. We assume that a finite gate voltage is applied
to the square-lattice tight-binding Hamiltonian of
Eq.~(\ref{eq:H-Sigma-0}) in such a way as to avoid the square-lattice
midband singularities:
\begin{equation}
  \label{eq:H-Sigma-0-bis}
  \hat{\cal H}_{g}= -W_g \sum_{k,\sigma_z}
  c_{k,\sigma_z}^+ c_{k,\sigma_z}
  .
\end{equation}

{\it The contacts between the normal and superconducting leads} are
captured by the following tight-binding Hamiltonian with hopping
amplitude $\Sigma^{(1)}$:
\begin{equation}
  \label{eq:H-Sigma}
  \hat{\cal H}_{\Sigma^{(1)}}=-\Sigma^{(1)} \sum_{\langle i',j' \rangle}
  \sum_{\sigma_z=\uparrow,\downarrow} \left(c_{i',\sigma_z}^+
  c_{j',\sigma_z}+ c_{j',\sigma_z}^+ c_{i',\sigma_z}\right)
  ,
\end{equation}
where $\sum_{\langle i',j' \rangle}$ runs over all tight-binding sites
on both sides of the contact.

{\it The magnetic field} is included by adding a phase to the hopping
amplitudes between the tight-binding sites $a$ and $b$:
\begin{equation}
  \Sigma_{a\rightarrow b} \rightarrow \Sigma_{a\rightarrow b}
  \exp\left(\frac{i e}{\hbar} \int_a^b {\bf A} \,.\, d{\bf s}\right) ,
\end{equation}
where ${\bf A}$ is the vector potential. In addition, the absence of
screening currents on the superconducting sides of the normal
metal-superconductor boundaries will be taken into account according
to the forthcoming subsection~\ref{sec:boundary}.

\subsection{Large-gap Hamiltonian at zero magnetic field}
\label{sec:large-gap}

In this subsection, we consider that the superconducting gaps are the
largest energy scales. This yields a large-gap Hamiltonian for the
entire device, which will afterwards be treated via exact
diagonalizations. The DC-Josephson currents are obtained from
numerically differentiating the ground state energy with respect to
the superconducting phase variable of the corresponding terminal.
Making the approximation of a large superconducting gap was developed
over the recent years, see for instance
Refs.~\onlinecite{Zazunov2003,Meng2009,Melin2021,Klees2020}.  Reaching
numerical efficiency for large-scale devices is the main motivation
for this large-gap limit.

{\it Large-gap Hamiltonian from wave-functions:} Now, we present a
wave-function calculation which yields the large-gap
Hamiltonian. Using generic compact matrix notations, the
starting-point Nambu Hamiltonian is expressed as the sum of three terms:

(i) {\it The infinite Nambu matrix of the superconducting tight-binding
  Hamiltonian $\hat{\cal H}_{S,S}$} is deduced from the BCS
Hamiltonian $\hat{\cal H}_{BCS}$ in
Eqs.~(\ref{eq:H-BCS1})-(\ref{eq:H-BCS2}). Those superconducting leads
are generically denoted as $S_1,\,...\,,S_n$ and $\hat{\cal H}_{S,S}$
is a matrix gathering all of the $\hat{\cal H}_{S_p,S_p}$, with
$p=1,\,...,\,n$.

In order to illustrate the discussion, we consider for simplicity that
the lead $S_p$ contains two tight-binding sites labeled by ``1'' and
``2'', which yields the following $4\times 4$ Nambu Hamiltonian
$\hat{\cal H}_{S_p,S_p}$:
\begin{equation}
  \label{eq:2x2}
  \hat{\cal H}_{S_p,S_p}=\left(\begin{array}{cccc}
    0 & \Delta_p e^{i\varphi_p} & -W_{1,2} & 0\\
    \Delta_p e^{-i\varphi_p} & 0 & 0 & W_{1,2}\\
    -W_{2,1} & 0 & 0 & \Delta_p e^{i\varphi_p}\\
    0 & W_{2,1} & \Delta_p e^{-i\varphi_p} & 0
  \end{array}\right)
  .
\end{equation}
With a three-site tight-binding cluster, we obtain the following
$6\times 6$ Nambu Hamiltonian:
\begin{eqnarray}
  \label{eq:3x3}
  &&  \hat{\cal H}_{S_p,S_p}=\\&&\left(\begin{array}{cccccc}
    0 & \Delta_p e^{i\varphi_p} & -W_{1,2} & 0 & -W_{1,3} & 0\\
    \Delta_p e^{-i\varphi_p} & 0 & 0 & W_{1,2} & 0 & W_{1,3}\\
    -W_{2,1} & 0 & 0 & \Delta_p e^{i\varphi_p} & -W_{2,3} & 0\\
    0 & W_{2,1} & \Delta_p e^{-i\varphi_p} & 0 & 0 & W_{2,3}\\
    -W_{3,1} & 0 & -W_{3,2} & 0 & 0 & \Delta_p e^{i\varphi_p} \\
    0 & W_{3,1} & 0 & W_{3,2} & \Delta_p e^{-i\varphi_p} & 0
  \end{array}\right)
  \nonumber
  ,
\end{eqnarray}
where the three tight-binding sites are labeled by $1,\,2$ and
$3$. The matrices in Eqs.~(\ref{eq:2x2}) and~(\ref{eq:3x3}) can be
extrapolated to an infinite number of tight-binding sites, also
taking the connectivity of the underlying lattice into
account. Finally, all of the $\hat{\cal H}_{S_p,S_p}$ are
concatenated into the global $\hat{\cal H}_{S,S}$ matrix.

(ii) {\it The finite Nambu matrix rectangular normal-metal
  tight-binding lattice Hamiltonian $\hat{\cal H}_{N_0,N_0}$} is
deduced from $\hat{\cal H}_{\Sigma^{(0)}}$ in Eq.~(\ref{eq:H-Sigma-0})
and $\hat{\cal H}_{g}$ in Eq.~(\ref{eq:H-Sigma-0-bis}).
The Nambu Hamiltonian $\hat{\cal H}_{N_0,N_0}$ takes the following
form for the two tight-binding sites labeled by $1$ and $2$:
\begin{equation}
  \label{eq:2x2-bis}
  \hat{\cal H}_{N_0,N_0}^{2\times 2}=\left(\begin{array}{cccc}
    W_g & 0 & -\Sigma^{(0)}_{1,2} & 0\\
    0 & -W_g & 0 & \Sigma^{(0)}_{1,2}\\
    -\Sigma^{(0)}_{2,1} & 0 & W_g & 0\\
    0 & \Sigma^{(0)}_{2,1} & 0 & -W_g
  \end{array}\right)
  .
\end{equation}
We obtain the following with the three tight-binding sites
labeled by $1$, $2$ and $3$:
\begin{eqnarray}
  \label{eq:3x3-bis}
  &&  \hat{\cal H}_{N_0,N_0}^{3\times 3}=\\&&\left(\begin{array}{cccccc}
    W_g & 0 & -\Sigma^{(0)}_{1,2} & 0 & -\Sigma^{(0)}_{1,3} & 0\\
    0 & -W_g & 0 & \Sigma^{(0)}_{1,2} & 0 & \Sigma^{(0)}_{1,3}\\
    -\Sigma^{(0)}_{2,1} & 0 & W_g & 0 & -\Sigma^{(0)}_{2,3} & 0\\
    0 & \Sigma^{(0)}_{2,1} & 0 & -W_g & 0 & \Sigma^{(0)}_{2,3}\\
    -\Sigma^{(0)}_{3,1} & 0 & -\Sigma^{(0)}_{3,2} & 0 & W_g & 0 \\
    0 & \Sigma^{(0)}_{3,1} & 0 & \Sigma^{(0)}_{3,2} &  0 & -W_g
  \end{array}\right)
  \nonumber
  ,
\end{eqnarray}
and Eqs.~(\ref{eq:2x2-bis})-(\ref{eq:3x3-bis}) are easily generalized
to an arbitrary number of entries.

(iii) {\it The finite Nambu matrix of the couplings $\hat{\cal
    H}_{N_0,S}$ and $\hat{\cal H}_{S,N_0}$ between the superconductors
  $S_p$ and the normal region $N_0$} is deduced from $\hat{\cal
  H}_{\Sigma^{(1)}}$ in Eq.~(\ref{eq:H-Sigma}).  The Nambu Hamiltonian
$\hat{\cal H}_{N_0,S_p}$ takes the following form with interfaces made
with the two tight-binding sites labeled by $1$ and $2$:
\begin{equation}
  \label{eq:2x2-ter}
  \hat{\cal H}_{N_0,S_p}^{2\times 2}=\left(\begin{array}{cccc}
    0 & 0 & -\Sigma^{(1)}_{1,2} & 0\\
    0 & 0 & 0 & \Sigma^{(1)}_{1,2}\\
    -\Sigma^{(1)}_{2,1} & 0 & 0 & 0\\
    0 & \Sigma^{(1)}_{2,1} & 0 & 0
  \end{array}\right)
  ,
\end{equation}
and we obtain the following for interfaces made with the three tight-binding sites
labeled by $1$, $2$ and $3$:
\begin{eqnarray}
  \label{eq:3x3-ter}
  &&  \hat{\cal H}_{N_0,S_p}^{3\times 3}=\\&&\left(\begin{array}{cccccc}
    0 & 0 & -\Sigma^{(1)}_{1,2} & 0 & -\Sigma^{(1)}_{1,3} & 0\\
    0 & 0 & 0 & \Sigma^{(1)}_{1,2} & 0 & \Sigma^{(1)}_{1,3}\\
    -\Sigma^{(1)}_{2,1} & 0 & 0 & 0 & -\Sigma^{(1)}_{2,3} & 0\\
    0 & \Sigma^{(1)}_{2,1} & 0 & 0 & 0 & \Sigma^{(1)}_{2,3}\\
    -\Sigma^{(1)}_{3,1} & 0 & -\Sigma^{(1)}_{3,2} & 0 & 0 & 0 \\
    0 & \Sigma^{(1)}_{3,1} & 0 & \Sigma^{(1)}_{3,2} &  0 & 0
  \end{array}\right)
  \nonumber
  ,
\end{eqnarray}
and the matrices appearing in
Eqs.~(\ref{eq:2x2-ter})-(\ref{eq:3x3-ter}) can be extended to an
arbitrary number of entries.

The components of the Bogoliubov-de Gennes wave-functions are denoted
as $\psi_{N_0}$ and $\psi_S$ for the normal conductor $N_0$ and the
$n$ superconducting leads $S_p$ respectively, with
$p=1,\,...,\,n$. Each of the $\psi_{N_0}$ and $\psi_S$ is defined
on the normal-metallic tight-binding graph $N_0$ and in all
tight-binding sites of each superconductor $S_p$.

The overall infinite Nambu Hamiltonian
$\hat{\cal H}$ takes the following matrix form:
\begin{equation}
  \hat{\cal H}=\left(\begin{array}{cc}
    \hat{\cal H}_{N_0,N_0} & \hat{\cal H}_{N_0,S}\\
    \hat{\cal H}_{S,N_0} & \hat{\cal H}_{S,S}
  \end{array}\right)
  .
\end{equation}
The Bogoliubov-de Gennes eigenvalue equation is defined as
\begin{equation}
  \label{eq:BDG}
  \hat{\cal H} \left(\begin{array}{c} \psi_{N_0}\\\psi_S \end{array}\right)
  = \omega \left(\begin{array}{c} \psi_{N_0}\\\psi_S \end{array}\right)
  ,
\end{equation}
where $\omega$ is the energy, and Eq.~(\ref{eq:BDG}) leads to the
following set of equations:
\begin{eqnarray}
  \label{eq:eq1}
  \hat{\cal H}_{{N_0},{N_0}} \psi_{N_0} + \hat{\cal H}_{{N_0},S} \psi_S &=& \omega \psi_{N_0}\\
  \hat{\cal H}_{S,{N_0}} \psi_{N_0} + \hat{\cal H}_{S,S} \psi_S &=& \omega \psi_S
  \label{eq:eq2}
  ,
\end{eqnarray}
where Eq.~(\ref{eq:eq1}) and Eq.~(\ref{eq:eq2}) contain a finite and
an infinite number of equations respectively. Eq.~(\ref{eq:eq2}) is
written as follows:
\begin{equation}
  \label{eq:psi-L}
  \psi_S= \left( \omega - \hat{\cal H}_{S,S} \right)^{-1} \hat{\cal H}_{S,{N_0}} \psi_{N_0}
  .
\end{equation}
Eq.~(\ref{eq:psi-L}) is now specialized to the Nambu components of
the superconducting Green's functions defined on the
superconducting side of the coupling Nambu Hamiltonians $\hat{\cal
  H}_{{N_0},S}$ and $\hat{\cal H}_{S,{N_0}}$. Then, inserting
Eq.~(\ref{eq:psi-L}) into Eq.~(\ref{eq:eq1}) leads to an eigenvalue
problem for a finite number of linear equations:
\begin{equation}
  \hat{\cal H}_{{N_0},{N_0}} \psi_{N_0} + \hat{\cal H}_{{N_0},S} \left( \omega - \hat{\cal H}_{S,S}
  \right)^{-1} \hat{\cal H}_{S,{N_0}} \psi_{N_0} = \omega \psi_{N_0}
  .
\end{equation}
This defines the effective self-energy $\hat{\Sigma}_{eff}(\omega)$
as
\begin{equation}
  \hat{\Sigma}_{eff}(\omega) \psi_{N_0} = \omega \psi_{N_0}
  ,
\end{equation}
with
\begin{eqnarray}
  \label{eq:Sigma-eff}
  \hat{\Sigma}_{eff}(\omega)&=&\hat{\cal H}_{{N_0},{N_0}}
  + \hat{\cal H}_{{N_0},S} \left( \omega - \hat{\cal H}_{S,S}
  \right)^{-1} \hat{\cal H}_{S,{N_0}}\\
  &=& \hat{\cal H}_{{N_0},{N_0}}
  + \hat{\Sigma}_{{N_0},S}^{(1)} \hat{g}_{S,S}(\omega) \hat{\Sigma}_{S,{N_0}}^{(1)}
  ,
  \label{eq:Sigma-eff-2}
\end{eqnarray}
where
\begin{equation}
  \hat{g}_{S,S}(\omega) =\left( \omega - \hat{\cal H}_{S,S}
  \right)^{-1}
\end{equation}
is the resolvent (i.e. the Green's function) of the infinite
superconducting leads and $\hat\Sigma_{{N_0},S}^{(1)}$ and
$\hat{\Sigma}_{S,{N_0}}^{(1)}$ are the Nambu hopping amplitudes in
$\hat{\cal H}_{{N_0},S}$ and $\hat{\cal H}_{S,{N_0}}$ respectively,
see also Eq.~(\ref{eq:H-Sigma}).

Up to this point, the superconducting gap was finite but now, we take
the limit of a large gap where $\hat{g}_{S,S}(\omega)$ becomes
independent on the energy $\omega$, i.e. $\hat{g}_{S,S}(\omega)\equiv
\hat{g}_{S,S}$ [see the forthcoming
  Eqs.~(\ref{eq:gll})-(\ref{eq:gll-infinite}) for the expression of
  the superconducting Green's functions.] The effective self-energy
$\hat{\Sigma}_{eff}(\omega)$ in
Eqs.~(\ref{eq:Sigma-eff})-(\ref{eq:Sigma-eff-2}) takes the form of the
following energy-independent effective Hamiltonian:
\begin{equation}
  \label{eq:H-eff-infini}
  \hat{\Sigma}_{eff}(\omega) \equiv\hat{\cal H}_{eff}=\hat{\cal
    H}_{{N_0},{N_0}} + \hat{\Sigma}_{{N_0},S}^{(1)} \hat{g}_{S,S} \hat{\Sigma}_{S,{N_0}}^{(1)} .
\end{equation}

{\it Large-gap Hamiltonian from Green's functions:} The large-gap
Hamiltonian given by Eq.~(\ref{eq:H-eff-infini}) can also be obtained
from the Dyson equations, see Ref.~\onlinecite{Melin2021}. Namely, the
fully dressed Green's function $\hat{G}_{{N_0},{N_0}}(\omega)$ at the
energy $\omega$ is calculated as follows:
\begin{eqnarray}
  \hat{G}_{{N_0},{N_0}}(\omega)&=&
  \hat{g}_{{N_0},{N_0}}(\omega)\\ \nonumber &&+
  \hat{g}_{{N_0},{N_0}}(\omega) \hat{\Sigma}_{{N_0},S}^{(1)}
  \hat{G}_{S,{N_0}}(\omega)\\
  \label{eq:Dyson-2}
  &=& \hat{g}_{{N_0},{N_0}}(\omega)
  \\&&+
  \nonumber
  \hat{g}_{{N_0},{N_0}}(\omega) \hat{\Sigma}_{{N_0},S}^{(1)}
  \hat{g}_{S,S}(\omega) \hat{\Sigma}_{S,{N_0}}^{(1)}
  \hat{G}_{{N_0},{N_0}}(\omega)
  .
\end{eqnarray}
Eq.~(\ref{eq:Dyson-2}) is written as
\begin{equation}
  \hat{G}_{{N_0},{N_0}}(\omega)=
  \left[\omega-
    \hat{\Sigma}_{eff}(\omega)\right]^{-1}
  ,
\end{equation}
where, in the large-gap approximation, the effective self-energy
$\hat{\Sigma}_{eff}(\omega)$ given by
Eqs.~(\ref{eq:Sigma-eff})-(\ref{eq:Sigma-eff-2}) takes the form of the
energy-$\omega$ independent Hamiltonian $\hat{\cal H}_{eff}$ given by
Eq.~(\ref{eq:H-eff-infini}), as it was obtained from this compact
Green's function calculation.

{\it Superconducting Green's functions:} Now, we provide the
expression of the superconducting Green's function $\hat{g}_{S_p,S_p}$
appearing in Eq.~(\ref{eq:H-eff-infini}), and we specifically
demonstrate that $\hat{g}_{S_p,S_p}(\omega)\equiv \hat{g}_{S_p,S_p}$
is independent on the energy $\omega$. The advanced local
superconducting Green's function of lead $S_p$ takes the following
form in the presence of a finite gap:
\begin{eqnarray}
  \label{eq:gll}
  && \hat{g}_{S_p,S_p}(\omega)=\\
  \nonumber
  && \frac{1}{W\sqrt{|\Delta|^2-(\omega-i\eta)^2}}
  \left(
  \begin{array}{cc}
    -\omega & |\Delta|e^{i\varphi_p}\\
    |\Delta|e^{-i\varphi_p} & -\omega
  \end{array}\right)
  ,
\end{eqnarray}
where $\eta$ is a small line-width broadening, i.e. the so-called
Dynes
parameter~\cite{Kaplan1976,Dynes1978,Pekola2010,Saira2012}. Eq.~(\ref{eq:gll}) can
be found in many papers. For instance, this Eq.~(\ref{eq:gll}) is the
starting point of the current-voltage characteristics calculations in
voltage-biased superconducting weak links \cite{Cuevas1996}.

The following is obtained in the large-gap approximation:
\begin{eqnarray}
  \label{eq:gll-infinite}
  \hat{g}_{S_p,S_p}&=&
  \frac{1}{W}
  \left(
  \begin{array}{cc}
    0 & e^{i\varphi_p}\\
    e^{-i\varphi_p} & 0
  \end{array}\right)
  ,
\end{eqnarray}
where Eq.~(\ref{eq:gll-infinite}) is energy-independent, as it was
anticipated in the above discussion. This Eq.~(\ref{eq:gll-infinite})
is next inserted into the expression Eq.~(\ref{eq:H-eff-infini}) of
the large-gap Hamiltonian, which is next numerically treated with
exact diagonalizations.

\subsection{Boundary conditions}
\label{sec:boundary}
In this subsection, we discuss how the large-gap Hamiltonian given by
Eq.~(\ref{eq:H-eff-infini}) is modified in the presence of a finite
value for the magnetic field applied perpendicularly to the
two-dimensional structure. In the presence of a vector potential ${\bf
  A}$, we make the substitution ${\bf p}\rightarrow {\bf p}+e{\bf
  A}$ for the momentum, and ${\bf j}\rightarrow (e\hbar/m) \left[{\bf
    \nabla} \varphi + (2 e/\hbar){\bf A}\right]$ for the
supercurrent ${\bf j}$, where $\varphi$ denotes the superconducting
phase variable. The vector potential is expressed in the gauge
$A_x=-By/2$ and $A_y=Bx/2$, where $B$ is the magnetic field.

Now, we calculate how a Cooper pair crosses the left contact from the
superconductor $S_L$ at coordinates $(x=x_L-a_0,\,y)$ to the
corresponding tight-binding site at $(x=x_L,\,y)$ in the normal
metal. Considering first the left superconductor, we implement
$\nabla_y \varphi +(2e/\hbar)A_y=0$ along the $S_L$-${N_0}$ interface,
leading to
\begin{equation}
  \varphi_y=-\frac{B(x_L-a_0) y}{\Phi'_0}+\varphi_L^{(0)}
  ,
\end{equation}
where $\Phi'_0=\hbar/e=\Phi_0/2\pi$, with $\Phi_0=h/e$ the
superconducting flux quantum. In a second step, we integrate the phase
gradient ${\bf \nabla} \varphi + (2 e/\hbar) {\bf A}$ in the
horizontal direction across the $S_L$-${N_0}$ interface:
\begin{equation}
  \int_{x_L}^{x_L-a_0} \left( {\bf \nabla} \varphi+\frac{2e}{\hbar}{\bf
    A}\right)\,.\,d {\bf s} = \frac{Bya_0}{\Phi'_0} +\varphi_y.
\end{equation}
Overall, we deduce the phase
\begin{equation}
  \varphi_L^{(0)}-\frac{Byx_L}{\Phi'_0} + \frac{2 B y a_0}{\Phi'_0}
  ,
\end{equation}
where $\varphi_L^{(0)}$ is the superconducting phase variable of the
left superconductor. The following self-energy is then included in the
normal-metal Hamiltonian on the left-hand-side of the rectangular
tight-binding lattice, i.e. at coordinate $(x=x_L,\,y)$:
\begin{equation}
  \label{eq:Gamma-Left}
  \Gamma_{loc}^{(Left)}(y)=-\frac{(\Sigma^{(1)})^2}{W}
  e^{i\varphi_L^{(0)}} e^{-iByx_L/\Phi'_0} e^{2 iB y a_0/\Phi'_0}
  ,
\end{equation}
where $\Gamma_{loc}^{(Left)}(y)$ denotes the electron-hole Nambu
component. Similarly, we deduce the following for the right, top and
bottom self-energies along the edges $x=x_R$, $y=y_T$ and $y=y_B$ of
the rectangle, respectively:
\begin{eqnarray}
  \label{eq:Gamma-Right}
  \Gamma_{loc}^{(Right)}(y)&=&-\frac{(\Sigma^{(1)})^2}{W}
  e^{i\varphi_R^{(0)}} e^{-iByx_R/\Phi'_0} e^{-2 iB y a_0/\Phi'_0}\\
  \label{eq:Gamma-Top}
  \Gamma_{loc}^{(Top)}(x)&=&-\frac{(\Sigma^{(1)})^2}{W}
  e^{i\varphi_T^{(0)}} e^{iBx y_T/\Phi'_0} e^{2 iB x a_0/\Phi'_0}\\
  \label{eq:Gamma-Bottom}
  \Gamma_{loc}^{(Bottom)}(x)&=&-\frac{(\Sigma^{(1)})^2}{W}
  e^{i\varphi_B^{(0)}} e^{iBxy_B/\Phi'_0} e^{-2 iB x a_0/\Phi'_0}
  .
\end{eqnarray}

%%%%%%%%%%%%%%%%%%%%%%%%%%%%%%%%%%%%%%%%%%%%%%%%%%%%%%%%%%%%%%%%%%%%%%
\begin{figure*}[t]
  \centerline{\includegraphics[width=.8\textwidth]{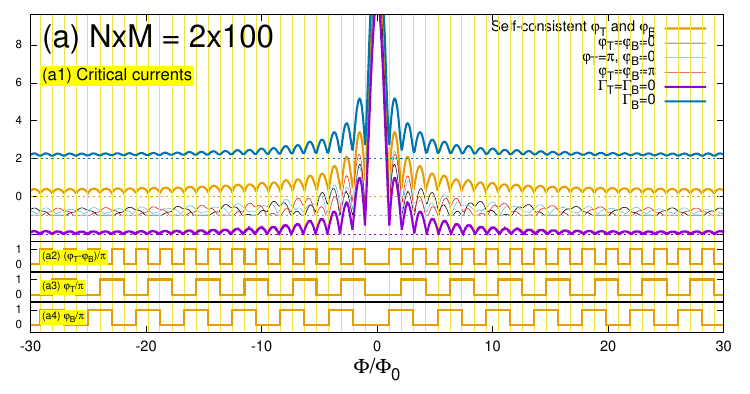}}
  \centerline{\includegraphics[width=.8\textwidth]{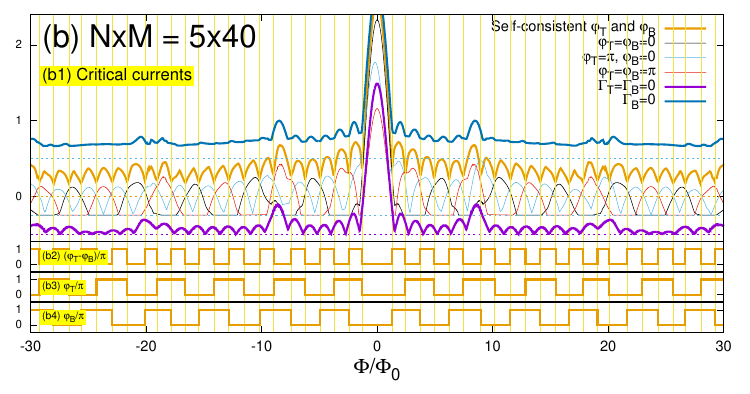}}
  \caption{The numerical results. The critical currents are shown as a
    function of the reduced magnetic flux, for the self-consistent
    solution (bold orange lines), and for the nonself-consistent
    $\varphi_T=\varphi_B=0$ (thin black lines) and $\varphi_T=\pi$,
    $\varphi_B=0$ (light blue lines). The thick magenta lines
    correspond to absence of coupling to the superconducting leads
    $S_T$ and $S_B$ on top and bottom, i.e. to a two-terminal
    Josephson junction with $\Sigma^{(1)}_B=\Sigma^{(1)}_T=0$. The
    thick blue lines show a three-terminal Josephson junction having
    an additional superconducting mirror, with
    $\Sigma^{(1)}_B=0$. Panels a2, a3 and a4 show the self-consistent
    $(\varphi_T-\varphi_B)/\pi$, $\varphi_T/\pi$ and $\varphi_B/\pi$
    respectively with two superconducting mirrors. We use
    $\Sigma_0=10$ for the bulk hopping amplitude in $N_0$, $\Gamma_L=
    \Gamma_R=\Gamma_T=\Gamma_B\equiv \Gamma$ with $\Gamma=1$ for the
    contact transparencies and $W_g=0.4$ for the value of the gate
    voltage. The supercurrents are in units of $2e \Gamma/\hbar$. We
    also use $N\times M=2\times 100$ (panel a) and $N\times M=5\times
    40$ (panel b). {Panel a shows oscillating critical
      current with $N\ll M$, i.e. with $N=2$ and $M=100$. Then, the
      magnetic oscillations resemble a Fraunhofer pattern. Panel b
      shows the evolution of the oscillating patterns for the smaller
      aspect ratio $N=5$ and $M=40$.}
    \label{fig:results-A}
  }
\end{figure*}
\begin{figure*}[t]
  \centerline{\includegraphics[width=.8\textwidth]{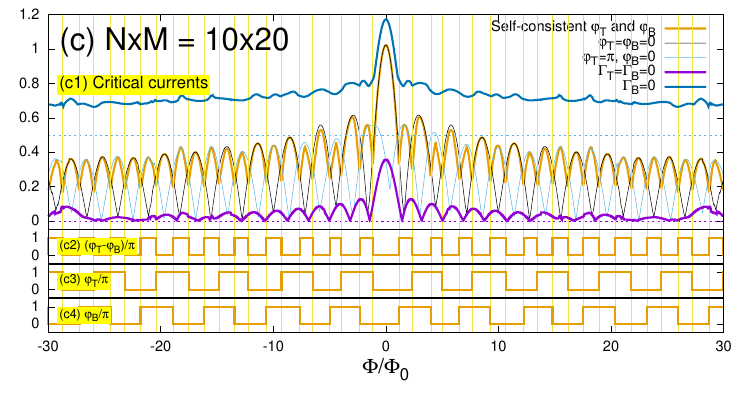}}
  \centerline{\includegraphics[width=.8\textwidth]{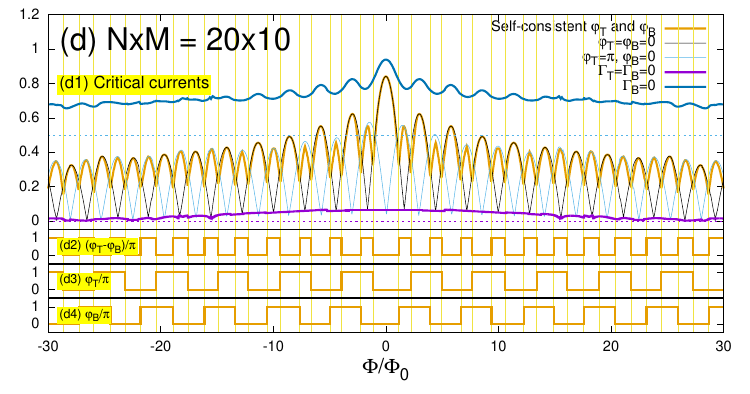}}
  \caption{The same as Fig.~\ref{fig:results-A} but now with $N\times
    M=10\times 20$ (panel c) and $N\times M=20\times 10$ (panel
    d). {Those panels c and d show the cross-over from
      {\it elongated along the $y$-axis direction} (panel c) to {\it
        elongated along the $x$-axis direction} (panel d). With two
      terminals, panel c shows an oscillation pattern while panel d
      features quasimonotonous decay of the critical current as a
      function of the magnetic field. In addition, the four-terminal
      critical current oscillation patterns resemble those a SQUID on
      panels c and d.}
    \label{fig:results-B}
  }
\end{figure*}
\begin{figure*}[t]
  \centerline{\includegraphics[width=.8\textwidth]{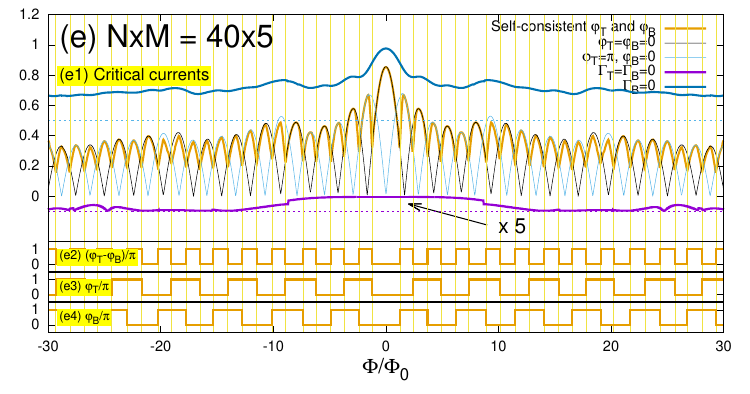}}
  \centerline{\includegraphics[width=.8\textwidth]{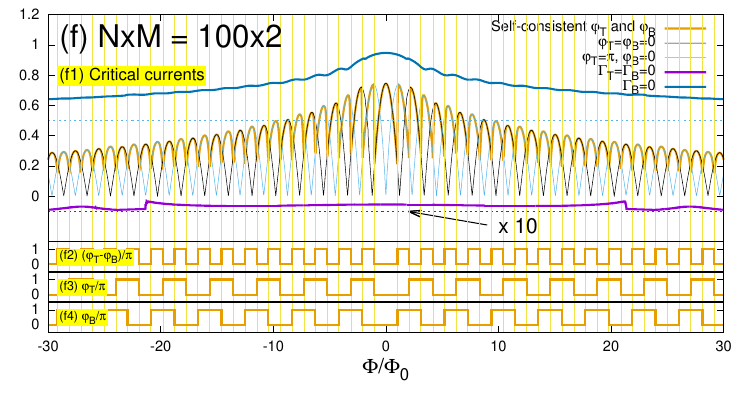}}
  \caption{The same as Fig.~\ref{fig:results-A} but in addition, the
    thin red lines correspond to $\varphi_T=\varphi_B=\pi$. We use
    $N\times M=40\times 5$ (panel e) and $N\times M=100\times 2$
    (panel f). {The figure shows aspect ratios
      strongly elongated along the $x$-axis direction, i.e. with $N\gg
      M$. Then, the two-terminal oscillation patterns reveal
      negligibly small signal, and the four-terminal ones show the
      SQUID-like oscillations coexisting with the long-range effect of
      the superconducting mirrors.}
    \label{fig:results-C}
  }
\end{figure*}
%%%%%%%%%%%%%%%%%%%%%%%%%%%%%%%%%%%%%%%%%%%%%%%%%%%%%%%%%%%%%%%%%%%%%

\subsection{Algorithm}
\label{sec:algo}

The numerical calculations proceed with exact diagonalizations of the
large-gap Hamiltonian defined in the above
subsections~\ref{sec:general}, \ref{sec:large-gap}
and~\ref{sec:boundary}. The supercurrents are obtained from the
derivative of the ground state energy with respect to the
superconducting phase variables. We denote by ${\cal E}_0\left(B,
\varphi_1,\,...,\,\varphi_n\right)$ the ground state energy:
\begin{eqnarray}
  \label{eq:E0}
  && {\cal E}_0\left( B,\varphi_1,\,...,\,\varphi_n\right)=\\&&\sum_\alpha
  \epsilon_\alpha\left( B,\varphi_1,\,...,\,\varphi_n\right)
  \theta\left[-\epsilon_\alpha\left(
    B,\varphi_1,\,...,\,\varphi_n\right)\right] ,
  \nonumber
\end{eqnarray}
where the ABS have the energies $\epsilon_\alpha\left(B,
\varphi_1,\,...,\,\varphi_n\right)$ and the Heaviside
$\theta$-function selects negative energies in the zero-temperature
limit. The current through lead $S_p$ is then given by
\begin{equation}
  I_{S_p}\left( B,\varphi_1,\,...,\,\varphi_n\right)
  =-\frac{2e}{\hbar} \frac{\partial {\cal E}_0}{\partial
    \varphi_p}\left( B,\varphi_1,\,...,\,\varphi_n\right)
  .
\end{equation}
{We next impose the constraint of vanishingly small
  supercurrent transmitted into the superconducting mirrors, and
  evaluate the critical current as the maximum over the remaining
  superconducting phase variables.}

\subsection{{Further physical remarks on the
    large-gap approximation}}

{We note that the large-gap approximation becomes
  exact only at low energy/long distance in highly-transparent
  superconductor-normal metal-superconductor junctions
  \cite{Kulik,Ishii,Bagwell}. As it is often the case in physics, we
  extend the large-gap calculations to all energy scales, not only
  considering the low energies at which the approximation is exact.}

{The coherence length $\xi_0$ in the large gap
  approximation is comparable to the Fermi wave-length $\lambda_F$,
  i.e. a few lattice spacings. The summation in Eq.~(\ref{eq:E0}) runs
  over the entire spectrum of ABS, thus addressing all the length
  scales in comparison with $\xi_0 \approx \lambda_F$.}

{The large-gap approximation fulfills the requirements
  of qualitatively capturing the supercurrent transmitted at long
  distance in the two-, three- or four-terminal configurations, as
  well as supercurrent lines between the lateral and the top or bottom
  superconductors transmitted over the short range $\xi_0\approx
  \lambda_F$ at the four corners of the normal-metallic rectangle. To
  summarize, we consider the large-gap approximation as an operational
  tool for capturing the qualitative behavior of those multiterminal
  Josephson junctions.}

\section{Results}
\label{sec:num}

In this section, we present and physically discuss the numerical
results obtained from the superconducting tight-binding model
presented in the above Sec.~\ref{sec:themodel}. Our main numerical
results are presented in Fig.~\ref{fig:results-A}a,
Fig.~\ref{fig:results-A}b, Fig.~\ref{fig:results-B}c,
Fig.~\ref{fig:results-B}d and Fig.~\ref{fig:results-C}e,
Fig.~\ref{fig:results-C}f, corresponding to the full range of the
aspect ratios. The corresponding device dimensions are $N\times
M=2\times 100,\,5\times 40,\, 10\times 20,\,20\times 10, \,40\times 5$
and $100\times 2$ respectively, with the fixed overall tight-binding
lattice area ${\cal S}=200\,a_0^2$.  The devices geometry ranges from
being elongated in the vertical to horizontal directions. The
presentation of the results may look unusual in the sense that the
discussion in the text proceeds with two terminals, next two terminals
plus a single superconducting mirror and finally two terminals plus
two superconducting mirrors, thus not consisting in a discussion of
the figures one after the other.

{Regarding the size of the numerically implemented
  rectangular lattices, we obtained a cross-over to the semiclassical
  spectra \cite{Kulik,Ishii,Bagwell} for larger dimensions, typically
  $100\times 200$ or $100 \times 400$ lattices (those data are not
  shown as figures in the present paper). However, the multiterminal
  effects that we consider do not rely on whether the semiclassical
  limit is fully realized. This is why we address here intermediate
  device dimensions at reduced computational expanses. The area is
  sufficient to produce viable numerical data for the critical current
  as a function of the magnetic field.}

Concerning the devices containing a single or two superconducting
mirrors, considerable gains in the computation times are obtained if
all of the superconducting leads $S_L$, $S_R$, $S_T$ and $S_B$ are
coupled to the normal-metallic conductor $N_0$ by symmetric hopping
amplitudes, see Appendix~\ref{sec:symmetries}. This symmetry condition
is fulfilled by the identical hopping amplitudes implemented in our
calculations.

After recovering known behavior with two terminals, the numerical
results with superconducting mirrors will next be presented and
discussed. The supercurrent flowing between the left and right
superconductors $S_L$ and $S_R$ in the horizontal direction will be
enhanced by orders of magnitudes in the presence of the single
superconducting mirror $S_T$. With the two superconducting mirrors
$S_T$ and $S_B$, we will obtain an oscillatory critical current
magnetic pattern that resembles the oscillations of a SQUID, due to
the interfering supercurrent paths through the top and bottom
superconductors $S_T$ and $S_B$.

{\it Two terminals:} Now, we proceed with discussing the numerical
results in themselves, starting with two terminals as a point of
comparison for testing the large-gap calculations. We first consider a
device where the two superconducting leads $S_L$ and $S_R$ are
connected to the left and right, without the superconducting mirror
$S_T$ and $S_B$, neither on top nor on bottom (see
Figs.~\ref{fig:the-device}a and~\ref{fig:the-device}b). The numerical
data with two terminals are shown with the bold magenta lines labeled
by $\Gamma_T=\Gamma_B=0$ on panels a1-f1 of Fig.~\ref{fig:results-A}
to Fig.~\ref{fig:results-C}.

Figs.~\ref{fig:results-A}a, Fig.~\ref{fig:results-A}b and
Fig.~\ref{fig:results-B}c correspond to $N\times M=2\times100$,
$N\times M=5\times 40$ and $N\times M=10\times 20$ respectively. We
then obtain the expected Fraunhofer-like oscillation pattern for those
devices elongated along the $y$-axis direction.

Next, the two-terminal critical current is negligibly small if the
device is elongated along the $x$-axis direction, see the bold magenta
lines labeled by $\Gamma_T=\Gamma_B=0$ in Fig.~\ref{fig:results-C}e
and Fig.~\ref{fig:results-C}f with $N\times M=40\times 5$ and $N\times
M=100\times 2$ respectively.

We also find quasimonotonous decay of the critical current as a
function of the magnetic field if the device dimension in the
horizontal direction is reduced according to $N\times M=20\times 10$,
see the bold magenta line on Fig.~\ref{fig:results-B}d. We carried out complementary
calculations of the ABS spectrum, revealing that the small ``jumps''
appearing in the datapoints represented by the bold magenta lines in
Fig.~\ref{fig:results-B}d signal that some ABS cross the zero of
energy as a function of the magnetic field.

The overall evolution from Fraunhofer pattern to quasimonotonous decay
of the critical current flowing from $S_L$ to $S_R$ is in a
qualitative agreement with a preceding work on disordered
superconductor-normal metal-superconductor junctions in a field, see
Ref.~\onlinecite{Cuevas2007}. Now that we demonstrated consistency
with known results, we further proceed with three- and
four-terminal devices containing a single or two superconducting
mirrors respectively.

{\it A single superconducting mirror:} Now, we consider that a third
superconducting lead $S_T$ is connected on top to the rectangular
normal-metallic conductor $N_0$, see Fig.~\ref{fig:the-device}c. We
calculate the maximal value of the supercurrent flowing between $S_L$
and $S_R$ connected to the left and right edges respectively. As
discussed above, $S_T$ on top is an open-circuit superconducting
mirror and the overall supercurrent transmitted into $S_T$ is
vanishingly small. However, $S_T$ can propagate supercurrent in the
direction parallel to its interface with $N_0$.

The corresponding data for the critical current in the presence of
this third superconducting mirror $S_T$ laterally connected on top are
shown by the dark blue lines labeled by $\Gamma_B=0$ in all
Fig.~\ref{fig:results-A}-a1 to Fig.~\ref{fig:results-C}-f1. Those
datapoints are vertically shifted according to the reference
represented by the horizontal blue dashed lines.

Devices elongated in the vertical direction produce oscillations in
the critical current as a function of the applied magnetic field, see
the dark blue lines in Fig.~\ref{fig:results-A}-a1 to
Fig.~\ref{fig:results-B}-d1 corresponding to $N\times M=2\times
100,\,5\times 40,\,10\times 20,\,20\times 10$ respectively. We note
that, for those device dimensions, the ratio between the critical
currents at the central peak and at the first lobe is anomalously
large in comparison with the standard Fraunhofer
pattern~\cite{Tinkham}. Given the intermediate contact transparencies
in our calculations, we possibly relate this zero-field anomaly to the
constructive interference of reflectionless tunneling at low magnetic
field, see Ref.~\onlinecite{Schechter2001}.

The corresponding critical currents flowing between the left and right
superconductors $S_L$ and $S_R$ in the horizontal direction are shown
by the dark blue lines labeled by $\Gamma_B=0$ on panels e1-f1 of
Fig.~\ref{fig:results-C}, for $N\times M=40\times 5$ and $N\times
M=100\times 2$. Those values are enhanced by orders or magnitude in
comparison with a two-terminal device (i.e. with $\Gamma_T=\Gamma_B=0$
in the absence of the coupling to $S_T$). This enhancement is
interpreted as phase rigidity in the superconductor mirror $S_T$
connected on top. Namely, propagating supercurrent from $S_L$ to $S_R$
in the horizontal direction involves supercurrent lines connecting
$S_L$ to $S_T$, followed by propagation over arbitrary long distances
inside the rigid condensate of $S_T$, and finally the supercurrent
lines are transmitted from $S_T$ to $S_R$.

{\it Two superconducting mirrors:} We now consider the four-terminal
Josephson device with two superconducting mirrors, where the
supercurrent in the horizontal direction flows between the two
superconductors $S_L$ and $S_R$ connected to the left and right edges
of the rectangular normal-metallic $N_0$, in the presence of the two
superconducting mirrors $S_T$ and $S_B$ laterally connected on top and
bottom, see Fig.~\ref{fig:the-device}d and Fig.~\ref{fig:the-device}e.

Panels a1-f1 of Fig.~\ref{fig:results-A}, Fig.~\ref{fig:results-B} and
Fig.~\ref{fig:results-C} show the critical currents as a function of
the magnetic field, with self-consistent superconducting phase
variables (see the bold orange lines labeled by ``Self-consistent
$\varphi_T$ and $\varphi_B$''). The self-consistent solution minimizes
the ground state energy ${\cal E}_0$ with respect to the
superconducting phase variables $\varphi_T,\,\varphi_B =0$ or $\pi$
according to Appendix~\ref{sec:symmetries}, see also Eq.~(\ref{eq:E0})
for the expression of the ground state energy ${\cal E}_0$.

As for a single superconducting mirror $S_T$, we observe that
connecting the two superconducting mirrors $S_T$ and $S_B$ on top and
bottom produces an enhancement of the critical current flowing between
the left and right superconductors $S_L$ and $S_R$ in the horizontal
direction, see Fig.~\ref{fig:results-C}a and Fig.~\ref{fig:results-C}b
for $N\times M=40\times 5$ and $N\times M=100\times 2$
respectively. The supercurrent from $S_L$ and $S_R$ or from $S_R$ to
$S_L$ in the horizontal direction can be viewed as being {\it guided}
by the superconducting mirrors $S_T$ and $S_B$ on top and bottom.

The critical current magnetic oscillations resemble those of a SQUID,
due to the interference between the Cooper pairs traveling in the
superconducting leads $S_T$ and $S_B$ on top and bottom respectively.

The thinner black lines labeled by $\varphi_T=\varphi_B=0$ in
Fig.~\ref{fig:results-A}a, Fig.~\ref{fig:results-A}b,
Fig.~\ref{fig:results-B}c, Fig.~\ref{fig:results-B}d,
Fig.~\ref{fig:results-C}e and Fig.~\ref{fig:results-C}f show the
critical current with the nonself-consistent $\varphi_T=\varphi_B=0$,
and the thinner light-blue lines labeled by ``$\varphi_T=\pi,\,
\varphi_B=0$'' correspond to the nonself-consistent $\varphi_T=\pi$ and
$\varphi_B=0$. The light-red lines labeled by
``$\varphi_T=\varphi_B=\pi$'' in Fig.~\ref{fig:results-A}a and
Fig.~\ref{fig:results-A}b correspond to $\varphi_T=\varphi_B=\pi$. We
conclude that the critical current calculated with the self-consistent
$\varphi_T$ and $\varphi_B$ (see the bold orange lines labeled by
``Self-consistent $\varphi_T$ and $\varphi_B$'') switches between
those nonself-consistent solutions as the magnetic field is increased.

Fig.~\ref{fig:results-A}-a2 to Fig.~\ref{fig:results-C}-f2 show the
normalized difference $(\varphi_T-\varphi_B)/\pi$ between the
self-consistent phase variables $\varphi_T$ and $\varphi_B$ of the
superconducting mirrors. Fig.~\ref{fig:results-A}-a3 to
Fig.~\ref{fig:results-C}-f3 and Fig.~\ref{fig:results-A}-a4 to
Fig.~\ref{fig:results-C}-f4 show the normalized self-consistent
$\varphi_T/\pi$ and $\varphi_B/\pi$ respectively. Remarkably, all
minima in the critical current pattern on panels a1-f1 correlate with
the magnetic field values at which $(\varphi_T-\varphi_B)/\pi$
switches between zero and unity or vice-versa. The thin vertical
yellow lines across each Fig.~\ref{fig:results-A} to
Fig.~\ref{fig:results-C} match all of those switching points in
$(\varphi_T-\varphi_B)/\pi$.

We conclude that, in the limit of a device elongated in the horizontal
direction, (i.e. with $N\gg M$), the magnetic field-dependence of the
critical current is controlled by $(\varphi_T-\varphi_B)/\pi$, instead
of each $\varphi_T/\pi$ or $\varphi_B/\pi$ taken individually. In the
opposite limit of a device elongated in the vertical direction
(i.e. if $M\gg N$), the superconducting phase variables $\varphi_T$
and $\varphi_B$ of $S_T$ and $S_B$ are {\it spectators}. Their values
is driven by the supercurrent flowing between $S_L$ and $S_R$ in the
horizontal direction. In addition, Fig.~\ref{fig:results-A}a and
Fig.~\ref{fig:results-A}b feature the magnetic flux dependence of the
nonself-consistent $\varphi_T=\varphi_B=\pi$, which strongly deviates
from the nonself-consistent $\varphi_T=\varphi_B=0$.

%%%%%%%%%%%%%%%%%%%%%%%%%%%%%%%%%%%%%%%%%%%%%%%%%%%%%%%%%%%%%%%%%%%%%%%
\begin{figure}[htb]
  \centerline{\includegraphics[width=.6\columnwidth]{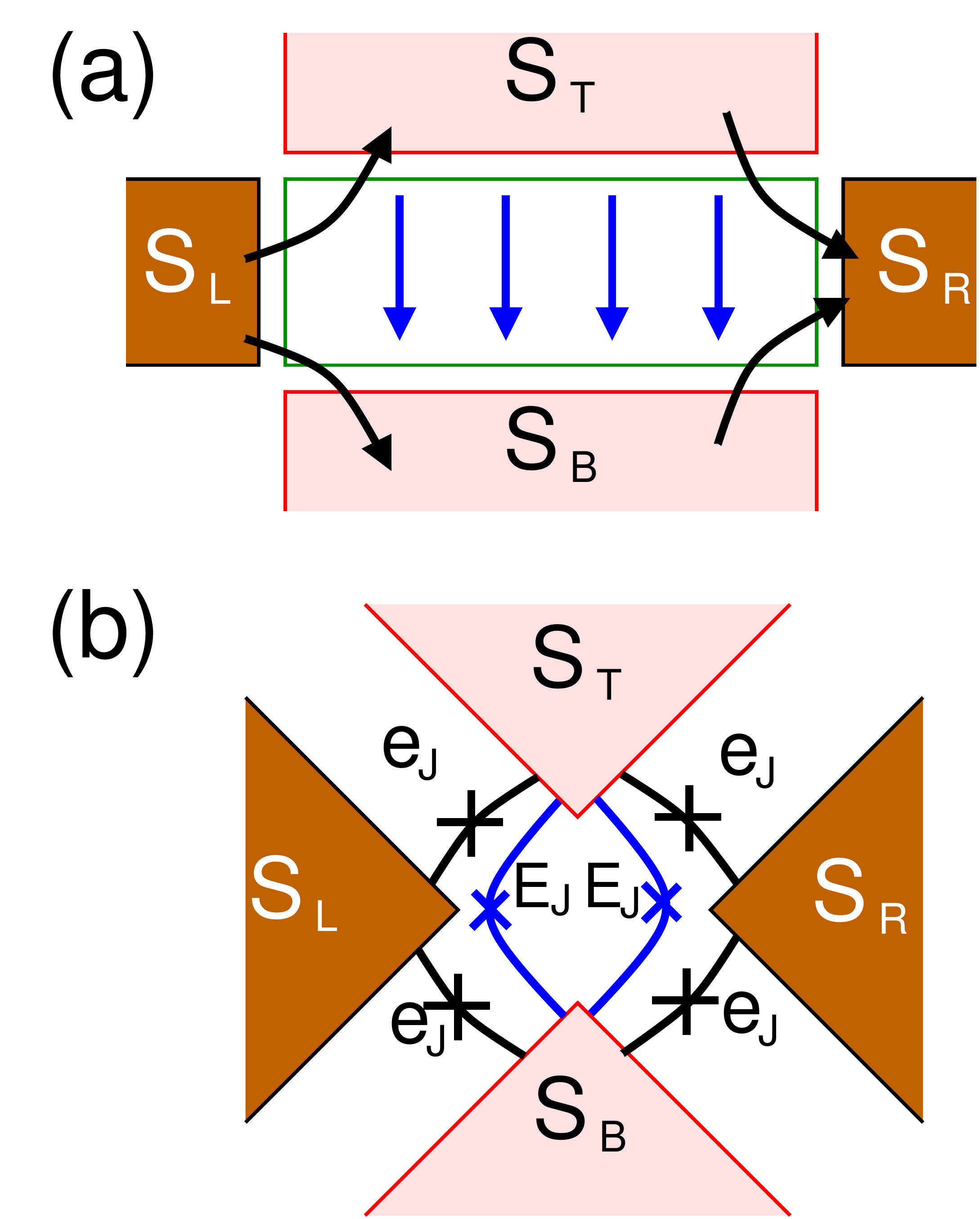}}
  \caption{The four superconducting leads $S_L$, $S_R$, $S_T$ and $S_B$
    on the left, right, top and bottom (a) are transformed into the
    phenomenological Josephson junction circuit model (b). The
    neighboring superconducting leads are connected by {\it small}
    Josephson coupling $e_J$ and the top and bottom ones $S_T$ and $S_B$
    are connected by two Josephson junctions with {\it large} Josephson
    coupling $E_J$, reflecting the corresponding large-area contacts
    between $S_T$ and $S_B$ through the normal metal ${N_0}$.
    \label{fig:phenomenological}
  }
\end{figure}
%%%%%%%%%%%%%%%%%%%%%%%%%%%%%%%%%%%%%%%%%%%%%%%%%%%%%%%%%%%%%%%%%%%%%%%

%%%%%%%%%%%%%%%%%%%%%%%%%%%%%%%%%%%%%%%%%%%%%%%%%%%%%%%%%%%%%%%%%%%%%%%
\begin{figure*}[htb]
  \centerline{\includegraphics[width=.7\textwidth]{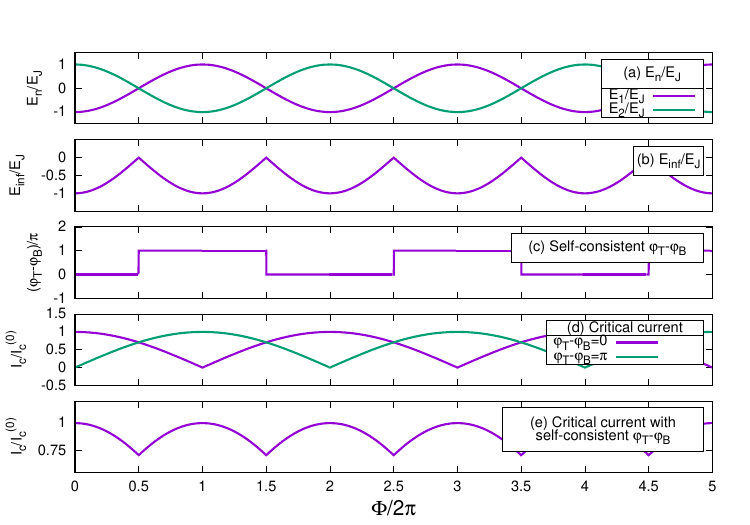}}
  \caption{The figure illustrates the phenomenological Josephson
    junction circuit model calculation. The figure shows the energies
    $E_1$ and $E_2$ as a function of $\Phi/2\pi$ [see Eq.~(\ref{eq:E1})]
    (a), the ground state energy $E_{inf}=\inf(E_1,E_2)$ between the
    lowest between $E_1$ and $E_2$ (b), the normalized self-consistent
    $(\varphi_T-\varphi_B)/2\pi$ (c), the nonself-consistent critical
    currents $I_c$ for $\varphi_T-\varphi_B=0$ and
    $\varphi_T-\varphi_B=\pi$ (d), and the critical current $I_c$ with
    the self-consistent $\varphi_T-\varphi_B$ (e).
    \label{fig:results-phenomenological}
  }
\end{figure*}
%%%%%%%%%%%%%%%%%%%%%%%%%%%%%%%%%%%%%%%%%%%%%%%%%%%%%%%%%%%%%%%%%%%%%%%

\section{Phenomenological Josephson junction circuit model}
\label{sec:phenomenological}

In this section, we propose a phenomenological Josephson junction
circuit model suitable to geometries elongated in the horizontal
direction, i.e. with $N\gg M$. The Josephson coupling energy
$E_J$ between the top and bottom superconductors $S_T$ and $S_B$ is
large, due to the corresponding large area interfaces. The Josephson
coupling energies $e_J$ between the pairs $(S_B,\,S_L)$,
$(S_L,\,S_T)$, $(S_T,\,S_R)$ and $(S_R,\,S_B)$ is smaller, see
Fig.~\ref{fig:phenomenological}. This simple model relies on a few
weak links and it is thus not intended to capture the zero-field
anomaly appearing in the above numerical calculations.

The total energy takes the form
\begin{eqnarray}
  E&=&-E_J
  \cos\left(\varphi_T-\varphi_B+\frac{\Phi}{2\Phi_0}\right)\\ \nonumber&&-
  E_J
  \cos\left(\varphi_T-\varphi_B-\frac{\Phi}{2\Phi_0}\right)\\ \nonumber
  &&-e_J \cos\left(\varphi_T-\varphi_L+\frac{\Phi}{4\Phi_0}\right)
  -e_J
  \cos\left(\varphi_L-\varphi_B+\frac{\Phi}{4\Phi_0}\right)\\ &&-e_J
  \cos\left(\varphi_B-\varphi_R+\frac{\Phi}{4\Phi_0}\right) -e_J
  \cos\left(\varphi_R-\varphi_T+\frac{\Phi}{4\Phi_0}\right)
  \nonumber
  .
\end{eqnarray}
Assuming $E_J\gg e_J$, the supercurrent entering $S_T$ is approximated
as
\begin{eqnarray}
  \label{eq:current-top}
  -\frac{2e}{\hbar} \frac{\partial E}{\partial \varphi_T} &\simeq& \frac{2e}{\hbar} E_J
  \sin\left(\varphi_T-\varphi_B+\frac{\Phi}{2\Phi_0}\right)\\ \nonumber
  &+& \frac{2e}{\hbar} E_J
  \sin\left(\varphi_T-\varphi_B-\frac{\Phi}{2\Phi_0}\right) .
\end{eqnarray}
Injecting $\varphi_T-\varphi_B=0$ or $\pi$ into
Eq.~(\ref{eq:current-top}) leads to the zero-current condition
$-(2e/\hbar) \partial E / \partial \varphi_T=0$, with the
corresponding energies $E_1$ and $E_2$
\begin{equation}
  \label{eq:E1}
  E_1=-2E_J \cos\left(\frac{\Phi}{2\Phi_0}\right) \equiv -E_2
\end{equation}
associated to $\varphi_T-\varphi_B=0$ and $\varphi_T-\varphi_B=\pi$
respectively. As the normalized magnetic flux $\Phi/\Phi_0$ increases,
the ground state energy alternates between $E_1$ and $E_2$ in
Eq.~(\ref{eq:E1}), corresponding to locking the phases $\varphi_T$ and
$\varphi_B$ according to $\varphi_T-\varphi_B=0$ or $\varphi_T -
\varphi_B=\pi$ respectively. For instance, $\varphi_T - \varphi_B=0$
and $\varphi_T - \varphi_B=\pi$ are obtained in the intervals
$|\Phi/2\Phi_0| < \pi/2$ and $\pi/2<|\Phi/2\Phi_0| < 3\pi/2$
respectively.

Fig.~\ref{fig:results-phenomenological}a to
Fig.~\ref{fig:results-phenomenological}c illustrate the
flux-sensitivity of $E_1,\,E_2$ in Eq.~(\ref{eq:E1}) (panel a), the
ground state energy $E_{inf}=\inf(E_1,\,E_2)$ (panel b) and the
self-consistent $\varphi_T - \varphi_B$ (panel c). Comments on panels
d and e are provided below.

Now, we successively evaluate the supercurrents for
$\varphi_T-\varphi_B=0$ and $\varphi_T-\varphi_B=\pi$. First
considering $\varphi_T = \varphi_B=0$ leads to the following
expression of the $\varphi_L$- and $\varphi_R$-sensitive energy terms
$E_L$ and $E_R$:
\begin{eqnarray}
  \nonumber
  && E_L(\varphi_L,\Phi)\\
  &=&-e_J \cos\left(-\varphi_L+\frac{\Phi}{4\Phi_0}\right) -e_J
  \cos\left(\varphi_L+\frac{\Phi}{4\Phi_0}\right)\\
  &=& -2 e_J \cos \varphi_L \cos\left(\frac{\Phi}{4\Phi_0}\right)\\
  \nonumber
  &&E_R(\varphi_R,\Phi)\\
  &=&-e_J
  \cos\left(-\varphi_R+\frac{\Phi}{4\Phi_0}\right) -e_J
  \cos\left(\varphi_R+\frac{\Phi}{4\Phi_0}\right)\\
  &=& -2 e_J \cos \varphi_R \cos\left(\frac{\Phi}{4\Phi_0}\right)
  .
\end{eqnarray}
We obtain
\begin{eqnarray}
  I_L(\varphi_L,\Phi)&=&-\frac{2e}{\hbar} \frac{\partial E_L}{\partial
    \varphi_L}(\varphi_L,\Phi)\\
  \label{eq:CPR1-1}
  &=&-\frac{4e}{\hbar} e_J \sin\varphi_L
  \cos\left(\frac{\Phi}{4\Phi_0}\right)\\ I_R(\varphi_R,\Phi)&=&-\frac{2e}{\hbar}
  \frac{\partial E_R}{\partial
    \varphi_R}(\varphi_R,\Phi)\\&=&-\frac{4e}{\hbar} e_J \sin\varphi_R
  \cos\left(\frac{\Phi}{4\Phi_0}\right)
  \label{eq:CPR1-2}
  .
\end{eqnarray}
The condition $I_R+I_L=0$ leads to $\sin \varphi_L=-\sin \varphi_R$,
and to $\varphi_R=-\varphi_L$ or $\varphi_R=\varphi_L+\pi$. We observe
that $E_L(\varphi_L,\Phi)+E_R(\varphi_L+\pi,\Phi)=0$, and we can
always find values of $\varphi_L$ having the lower energy
$E_L(\varphi_L,\Phi)+E_R(-\varphi_L,\Phi)<0$, which is why we restrict
to $\varphi_R = -\varphi_L \equiv \psi$. It turns out that the ground
state energy is negative for all values of the reduced magnetic flux
$\Phi/\Phi_0$, see Fig.~\ref{fig:results-phenomenological}b.

Assuming now $\varphi_T=0$ and $\varphi_B=\pi$, we obtain
\begin{eqnarray}
  && E'_L(\varphi_L,\Phi)\nonumber\\ &=& -e_J
  \cos\left(-\varphi_L+\frac{\Phi}{4\Phi_0}\right) +e_J
  \cos\left(\varphi_L+\frac{\Phi}{4\Phi_0}\right)\\ &=& -2e_J \sin
  \varphi_L
  \sin\left(\frac{\Phi}{4\Phi_0}\right)\\ &&E'_R(\varphi_R,\Phi)\nonumber\\&=&e_J
  \cos\left(-\varphi_R+\frac{\Phi}{4\Phi_0}\right) -e_J
  \cos\left(\varphi_R+\frac{\Phi}{4\Phi_0}\right)\\ &=& 2e_J \sin
  \varphi_R \sin\left(\frac{\Phi}{4\Phi_0}\right) ,
\end{eqnarray}
and
\begin{eqnarray}
  I'_L(\varphi_L,\Phi)&=&-\frac{2e}{\hbar} \frac{\partial
    E'_L}{\partial \varphi_L}(\varphi_L,\Phi) \\\label{eq:CPR2-1}
  &=&\frac{4e}{\hbar} e_J \cos\varphi_L
  \sin\left(\frac{\Phi}{4\Phi_0}\right)\\ I'_R(\varphi_R,\Phi)&=&-\frac{2e}{\hbar}
  \frac{\partial E'_R}{\partial
    \varphi_R}(\varphi_R,\Phi)\\ &=&-\frac{4e}{\hbar} e_J
  \cos\varphi_R \sin\left(\frac{\Phi}{4\Phi_0}\right)
  \label{eq:CPR2-2}
  ,
\end{eqnarray}
where, again, we used $\varphi_R=-\varphi_L\equiv \psi$.

Fig.~\ref{fig:results-phenomenological}d shows the critical current as
a function of the normalized magnetic flux for the nonself-consistent
solutions with $\varphi_T-\varphi_B=0$ and $\varphi_T-\varphi_B=\pi$.
Fig.~\ref{fig:results-phenomenological}e shows the value of the
supercurrent calculated with the self-consistent
$\varphi_T-\varphi_B$, which amounts to taking the maximum between the
two values on Fig.~\ref{fig:results-phenomenological}d. We note
consistency with the preceding numerical calculations presented in
Fig.~\ref{fig:results-A} to Fig.~\ref{fig:results-C}, see the above
Sec.~\ref{sec:num}.

Finally, we have four phase variables
$\varphi_L,\,\varphi_R,\,\varphi_T$ and $\varphi_B$. The constraint
$\varphi_R = -\varphi_L \equiv \psi$ originates from the external
current source which imposes opposite supercurrents transmitted into
$S_L$ and $S_R$, therefore defining a net current flowing from $S_L$
to $S_R$ or from $S_R$ to $S_L$ in the horizontal direction. Those
opposite supercurrents $I_R=-I_L$ couple to the remaining phase
combinations $\varphi_T-\varphi_B=0$ or $\pi$ and $\varphi_R =
-\varphi_L \equiv \psi$, where $\varphi_B$ is left undetermined. This
is compatible with gauge invariance where one of those superconducting
phase variables cannot be fixed. We conclude {that, if
  $N\gg M$, the} supercurrent flowing from $S_L$ to $S_R$ or from
$S_R$ to $S_L$ in the horizontal direction couples to all possibly
allowed phase combinations, as it is already the case in the
short-junction limit.

\section{Conclusions}
\label{sec:conclusions}

To conclude, we considered a multiterminal Josephson junction circuit
model with the four superconducting leads $S_L$, $S_R$, $S_T$ and
$S_B$ connected to the left, right, top and bottom edges of a
normal-metallic rectangle $N_0$.

Concerning three terminals, we demonstrated that, for devices
elongated in the horizontal direction, attaching the superconducting
mirror $S_T$ on top of the normal conductor $N_0$ enhances the
horizontal supercurrent by orders of magnitude, as a result of
phase rigidity in the open-circuit superconductor~$S_T$.

Concerning four terminals, we calculated the supercurrent flowing from
$S_L$ to $S_R$ in the horizontal direction in the presence of
the two superconducting mirrors $S_T$ and $S_B$, and we obtained
oscillatory magnetic oscillations reminiscent of a SQUID. Those
oscillations are controlled by the self-consistent phase variables
$\varphi_T$ and $\varphi_B$ of the superconductors $S_T$ and $S_B$
connected on top and bottom respectively.

If the hopping amplitudes connecting the ballistic rectangular
normal-metallic conductor $N_0$ to the superconductors are symmetric,
then $\varphi_T$ and $\varphi_B$ take the values $0$ or $\pi$, as for
an emerging Ising degree of freedom.

We also interpreted our numerical results with a simple Josephson
junction circuit model, and demonstrated that the supercurrent flows
through all parts of the circuit if the device is elongated in the
horizontal direction.

In the numerical calculations and in the phenomenological circuit
model, the horizontal supercurrent was controlled by the difference
$\varphi_T-\varphi_B=0$ or $\pi$ instead of each individual
$\varphi_T$ or $\varphi_B$, thus providing sensitivity to a single
effective Ising degree of freedom of the supercurrent flowing in the
horizontal direction.

{Finally, a long-range effect was reported in the
  experimental Ref.~\onlinecite{Zhang2022}, which is compatible with
  our theory of the superconducting mirrors. In addition,} a recent
experimental work~\cite{Pankratova2020} measured the critical current
contours (CCCs) in the plane of the two biasing currents $I_1$ and
$I_2$. The zero-current conditions $I_1=0$ or $I_2=0$ are fulfilled at
the points where the CCCs intersect the $x$- or $y$-current axis
respectively. Thus, our theory of the phase rigidity is expected to
produce specific signatures on the CCCs, which will be the subject of
a future work. {Perspectives also include
  generalization to Josephson junction arrays
  \cite{Delsing1,Delsing2,Duty,Draelos2019,Arnault2021}.}

\section*{Acknowledgements}

The authors benefited from fruitful discussions with M. d'Astuto,
D. Beckmann, J.G. Caputo, H. Cercellier, I. Gornyi, T. Klein,
F. L\'evy-Bertrand, M.A. M\'easson, P. Rodi\`ere. R.M. thanks the
Infrastructure de Calcul Intensif et de Donn\'ees (GRICAD) for use of
the resources of the M\'esocentre de Calcul Intensif de l’Universit\'e
Grenoble-Alpes (CIMENT). This work was supported by the International
Research Project SUPRADEVMAT between CNRS in Grenoble and KIT in
Karlsruhe. This work received support from the French National
Research Agency (ANR) in the framework of the Graphmon project
(ANR-19-CE47-0007). This work was partly supported by Helmholtz
Society through program NACIP and the DFG via the Project No. DA
1280/7-1.

\appendix

\section{Symmetries}
\label{sec:symmetries}

In this Appendix, we show how the symmetries considerably reduce the
computation times if the current flowing from $S_L$ to $S_R$ in the
horizontal direction is specifically evaluated. Namely, we
demonstrate that the following symmetries:
\begin{equation}
  \label{eq:condition}
  \varphi_T,\,\varphi_B =0\mbox{ or }\pi \mbox{, and }
  \varphi_L=-\varphi_R\equiv\psi
\end{equation}
are equivalent to vanishingly small supercurrent transmitted into the
top and bottom superconductors $S_T$ and $S_B$,
i.e. (\ref{eq:condition}) implies that $S_T$ and $S_B$ are
superconducting mirrors. The condition (\ref{eq:condition}) also
implies that opposite supercurrents are transmitted into $S_L$ and
$S_R$ connected on the left and right edges of the rectangular
normal-metallic conductor $N_0$. Conservation of the supercurrent
between $S_L$ and $S_R$ in the horizontal direction is thus
automatically fulfilled. Now, we demonstrate those statements.

Eqs.~(\ref{eq:Gamma-Left})-(\ref{eq:Gamma-Right}) become
\begin{eqnarray}
  \Gamma_{loc}^{(Left)}(y)&=&-\frac{(\Sigma^{(1)})^2}{W}
  e^{i\varphi_L^{(0)}} e^{iB L y/2\Phi'_0} e^{2 iB y a_0/\Phi'_0}\\
  \Gamma_{loc}^{(Right)}(y)&=&-\frac{(\Sigma^{(1)})^2}{W}
  e^{i\varphi_R^{(0)}} e^{-iBL y/2\Phi'_0} e^{-2 iB y a_0/\Phi'_0}
  ,
\end{eqnarray}
where we use the notation $x_{R/L}=\pm L/2$. We obtain
\begin{equation}
  \Gamma_{loc}^{(Left)}(y)=\left(\Gamma_{loc}^{(Right)}\right)^*(y)
\end{equation}
if $e^{i\varphi_L^{(0)}}=e^{-i\varphi_R^{(0)}}$, i.e. if $\varphi_L =
-\varphi_R \equiv \psi $, see the condition~(\ref{eq:condition}).

Conversely, the substitution $x\rightarrow \overset{\sim}{{x}}=-x$ leads to
$\Gamma_{loc}^{(Top)} \rightarrow \overset{\sim}{{\Gamma}}_{loc}^{(Top)}$ and
$\Gamma_{loc}^{(Bottom)} \rightarrow \overset{\sim}{{\Gamma}}_{loc}^{(Bottom)}$
in Eqs.~(\ref{eq:Gamma-Top})-(\ref{eq:Gamma-Bottom}), with
\begin{eqnarray}
  \overset{\sim}{{\Gamma}}_{loc}^{(Top)}(x)&=&-\frac{(\Sigma^{(1)})^2}{W}
  e^{i\varphi_T^{(0)}} e^{-iBWx/2\Phi'_0} e^{-2 iB x
    a_0/\Phi'_0}\\ \overset{\sim}{{\Gamma}}_{loc}^{(Bottom)}(x)&=&-\frac{(\Sigma^{(1)})^2}{W}
  e^{i\varphi_B^{(0)}} e^{iBWx/2\Phi'_0} e^{2 iB x a_0/\Phi'_0},
\end{eqnarray}
where we used the notation $y_{T,B}=\pm W/2$.

We deduce the following:
\begin{eqnarray}
  \overset{\sim}{{\Gamma}}_{loc}^{(Top)} &=&
  \left({\Gamma}_{loc}^{(Top)}\right)^*\\ \overset{\sim}{{\Gamma}}_{loc}^{(Bottom)}
  &=& \left({\Gamma}_{loc}^{(Bottom)} \right)^*
\end{eqnarray}
if both $e^{i\varphi_B^{(0)}}$ and $e^{i\varphi_T^{(0)}}$ are
real-valued, i.e. if
$\varphi_B^{(0)},\,\varphi_T^{(0)}=0$ or $\pi$, see the
condition~(\ref{eq:condition}).

Now, we discuss the consequences for the supercurrents flowing across
the normal-metallic conductor $N_0$. At the lowest order in tunneling,
the typical combinations
\begin{equation}
  \Gamma_{loc}^{(Left)}(y) \left({\Gamma}_{loc}^{(Bottom)}(x)\right)^*
\end{equation}
and
\begin{equation}
  \Gamma_{loc}^{(Right)}(y)
  \left({\Gamma}_{loc}^{(Bottom)}(x)\right)^*
\end{equation}
control the DC-Josephson effect between the left/bottom and the
right/bottom superconducting leads. The following identity:
\begin{eqnarray}
  &&\Gamma_{loc}^{(Left)}(y) \left({\Gamma}_{loc}^{(Bottom)}(x)\right)^*\\
  &=&
  \left[\Gamma_{loc}^{(Right)}(y) \left({\Gamma}_{loc}^{(Bottom)}(-x)\right)^*\right]^*
  \nonumber
\end{eqnarray}
leads to opposite values for the supercurrents transmitted from left
to bottom and from right to bottom if the
condition~(\ref{eq:condition}) is fulfilled, since the corresponding
superconducting phase differences are opposite.

We conclude that the mirror-axis symmetry $x\rightarrow
\overset{\sim}{{x}}=-x$ leads to vanishingly small value for the sum
$I_{L\rightarrow B} + I_{R\rightarrow B}$ of the supercurrents
$I_{L\rightarrow B}$ (from left to bottom) and $I_{R\rightarrow B}$
(from right to bottom), i.e. $I_{L\rightarrow B} + I_{R\rightarrow
  B}=0$. Similarly, we find $I_{L\rightarrow T}+I_{R\rightarrow T}=0$
for the sum of the supercurrents from left to top and from right to
top.

Using the form of the Bethe-Salpeter equations suitable to Andreev
tubes (see for instance Refs.\onlinecite{Kraft2018,Meier2016}
for the Andreev tubes), this perturbative argument can be extended to
all orders in the tunneling amplitudes $\Sigma^{(1)}$ connecting the
normal region ${N_0}$ to each of the superconducting leads, see
Eq.~(\ref{eq:H-Sigma}) for the notation $\Sigma^{(1)}$.

In Sec.~\ref{sec:num} of the main text, the three- and four-terminal
calculations with a single or two superconducting mirrors respectively
are realized with identical value for all of the tunneling amplitudes
between the normal region $N_0$ and the superconductors. The symmetry
condition~(\ref{eq:condition}) is then automatically fulfilled and the
energy minimum is within the discrete set $\varphi_T,\,\varphi_B=0$ or
$\pi$. Scanning those restricted values of $\varphi_T$ and $\varphi_B$
(as it was the case in the above Sec.~\ref{sec:num}) allows for
considerable gain in the computation time with respect to looking for
the energy minimum in the entire $\left[0,\,2\pi\right]\times
\left[0,\,2\pi\right]$ intervals.

\end{document}